\documentclass[namedreferences]{solarphysics}

\usepackage[hyperref,optionalrh,showbiblabels]{spr-sola-addons} % For Solar Physics
\usepackage{graphicx}        % For eps figures, newer & more powerfull
\usepackage{color}           % For color text: \color command
\newcommand{\aap}{    {\it Astron. Astrophys.}}

\newcommand{\apj}{    {\it Astrophys. J.}}
\newcommand{\apjl}{   {\it Astrophys. J. Lett.}}

\newcommand{\mnras}{  {\it Mon. Not. Roy. Astron. Soc.}}

\newcommand{\solphys}{{\it Solar Phys.}}

\newcommand{\ssr}{    {\it Space Sci. Rev.}}
\chardef\us=`\_

\begin{document}

\begin{article}
\begin{opening}

\title{Evolution of the Sun's Polar-fields and the Poleward Transport of
Remnant Magnetic Flux
}

\author[addressref={aff1},corref,email={e-mail.avm@iszf.irk.ru}]{\inits{A.V.}\fnm{A.V. }~\lnm{Mordvinov}\orcid{}}%\sep
\author[addressref={aff1,aff2},email={e-mail.kit@iszf.irk.ru}]{\inits{L.L.}\fnm{L.L.}~\lnm{Kitchatinov}}%\sep
\address[id=aff1]{Institute of Solar-Terrestrial Physics of Siberian Branch of Russian Academy of Sciences, Lermontov st., 126a, Irkutsk 664033, Russia}
\address[id=aff2]{Pulkovo Astronomical Observatory, Russian Academy of Sciences, Pulkovskoe sh. 65,St. Petersburg, 196140, Russia}

\runningauthor{A.V.~Mordvinov, L.L.~Kitchatinov }
\runningtitle{Evolution of the Sun's Polar-fields}

\begin{abstract}
Synoptic magnetograms and relevant proxy data were analyzed to study the evolution of the Sun's polar magnetic fields. Time-latitude analysis of large-scale magnetic fields demonstrates cyclic changes in their zonal structure and the polar-field buildup. The time-latitude distributions of the emergent and remnant magnetic flux enable us to examine individual features of recent cycles. The poleward transport of predominantly following polarities contributed much of the polar flux and led to polar-field reversals. Multiple reversals of dominant polarities at the Sun's poles were identified in Cycles 20 and 21. Three-fold reversals were caused by remnant flux surges of following and leading polarities.
Time-latitude analysis of solar magnetic fields in Cycles 20--24 revealed zones which are characterized by a predominance of negative (non-Joy's) tilts
and appearance of active regions which violate Hale's polarity law.
The decay of non-Joy's and anti-Hale's
active regions result in remnant flux surges which disturb the usual order in magnetic flux transport and sometimes lead to multiple reversals of polar fields.
The analysis of local and large-scale magnetic fields in their causal relation improved our understanding of the Sun's polar field weakening.
\end{abstract}
\end{opening}

\section{Introduction}
     \label{S-Int}
Long-term measurements of solar magnetic fields revealed a significant weakening of the Sun's polar fields \citep{Hoeksema10, Petrie15}. During recent decades, sunspot activity levels have also decreased systematically. In Cycle 24, magnetic activity was characterized by a significant north-south asymmetry and asynchronous reversal of the Sun's polar fields \citep{Sun15, Golubeva16, Golubeva17}. This unusual behavior of solar activity is a challenge to the dynamo theory.

The flux-transport dynamo models initiated by \cite{Wang91} and first developed by \citep{Durney95, Choudhuri95} generally reproduce the main features of  polar field buildup and reversals. Incorporating data on the magnetic fluxes of active regions (ARs) and their tilts as initial conditions, the models reproduce time-spatial patterns of large-scale magnetic fields in Cycles 15--21 \citep{Baumann04, Cameron18, Jiang13, Jiang14}.

The latitudinal dependence of AR tilts relative to the equator (Joy's law) is a basic property of emergent magnetic flux \citep{Kosovichev08,  Pevtsov14, McClintock16}. The predominance of positive-tilt ARs was the key ingredient in the Babcock-Leighton dynamo \citep{Babcock61, Leighton69} and remains in its modern extensions \citep{Choudhuri12, Olemskoy13, Munoz13}. Analysis of  sunspot data from Mount Wilson and Kodaikanal observatories revealed changes in sunspot group tilts from cycle to cycle \citep{Erofeev04, Dasi10, Dasi13, McClintock13}.

 The polar field buildup mainly depends on appearance of emergent flux,  AR tilts and the speed of the meridional flows.
Nevertheless, flux transport modelling could not reproduce weak polar field in Cycles 23, 24 \citep{Jiang10, Upton14}.
Remnant flux of following polarities contribute most of the polar flux. However, remnant flux surges of opposite polarities sometimes also occur.
In terms of Hale's law, these surges usually correspond to dominant leading polarities in every hemisphere.

 Joy's law recognizes the latitudinal dependence of AR tilt as a weak tendency with a significant scatter \citep{Pevtsov14, McClintock16, Tlatova18}. In many of the bipolar magnetic groups, leading-polarity spots appear at higher latitudes than the following-polarity spots.
The decay of negative-tilt ARs results in  leading-polarity unipolar magnetic regions (UMRs) at higher latitudes as compared to those for following ones.
 Starting
from higher latitudes, leading-polarity UMRs  are usually  transported poleward.

The poleward transport of  abnormal  UMRs makes related surges of leading-polarity  \citep{Yeates15, Mordvinov15, Mordvinov16, Lockwood17}. Leading-polarity surges are usually associated with large activity complexes nearby or within the domains of negative tilt predominance. Leading-polarity surges result in cancellation of opposite magnetic polarities in the Sun's polar zones and the polar-field weakening in Cycles 23--24 \citep{Kitchatinov18}.

 Sometimes peculiar ARs appear whose orientation violates Hale's polarity law \citep{Li12}.
These ARs play an important role in the Sun's polar field evolution \citep{Stenflo12}.
There are empirical and theoretical reasons to expect that ARs of opposite (anti-Hale) orientation also contribute to the polar field
weakening \citep{Jiang15}.
The polar effect of anti-Hale's ARs could be estimated taking into account their exact orientation \citep{McClintock16, Li18}.

We analyzed synoptic magnetograms and essential properties of magnetic bipolar regions
to study the Sun's polar-field buildup in Cycles 20--24. Time-latitude analysis of large-scale magnetic fields enables us to study individual features of recent cycles.
 Taking into account the time-latitude distribution of emergent magnetic flux and its peculiar properties we interpret  multiple polar-field reversals
in Cycles 20 and 21.

\section{Evolution of the Sun's Magnetic Fields in Cycles 21--24}
\label{S-Evol}

\begin{figure}    %%%%%%%%%%%%%%%%%% FIGURE 1
   \centerline{\includegraphics[width=0.95\textwidth,clip=]{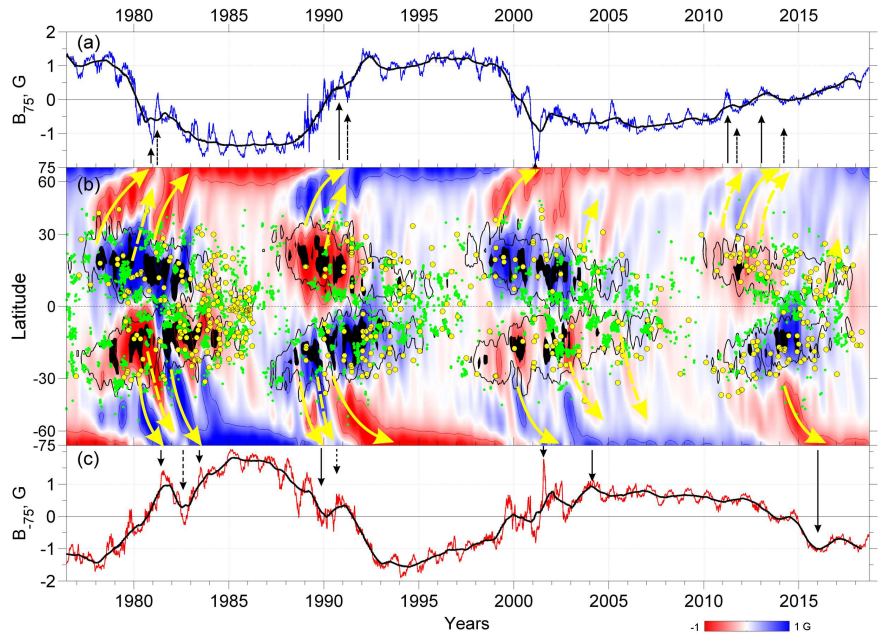}
              }

\caption{Changes in the magnetic fluxes at the heliographic latitude $\pm$75$^\circ$ in the northern (a) and southern (c) hemispheres derived from the WSO synoptic maps. The original and denoised values of the polar fluxes are shown in blue/red and black, respectively. The solid/dashed vertical arrows depict the  effects of remnant-flux surges for following/leading polarities. Zonally averaged magnetic fields are shown in red-to-blue (b). Zones of intense sunspot activity, domains of negative tilt and  anti-Hale's ARs are shown in black,  green, and yellow, respectively. }
   \label{F-TL}
   \end{figure}

The photospheric synoptic maps from Wilcox Solar Observatory characterize cyclic and long-term changes in the Sun's magnetic fields on a stable native scale for 1976--2018 \citep{Duvall77}. We studied changes in the radial component magnetic field ($B$) derived from the WSO synoptic maps. Magnetic flux values at  latitude $\pm$75$^\circ$ were assembled in chronological order. The magnetic fluxes in the northern/southern hemispheres are shown in blue/red
(Figures~\ref{F-TL}a,c). Such an approach presents the high-latitude magnetic fields at a cadence of about 9 hours. The time series quantify the polar-field buildup in details. We smoothed the time series using algorithm based on the
non-decimated wavelet transform \citep{Starck06}.
The denoised timeseries of magnetic flux ($B_{\pm±75}$) are shown in black (Figures~\ref{F-TL}a,c). Figures~\ref{F-TL}a,c show cyclic patterns in the Sun's magnetic field and the polar field reversals. Annual variations appear in high-latitude magnetic fields due to the Earth's excursions relative to the helioequator.

We analyzed changes in high-latitude magnetic fields to understand the polar-field buildup in relation to the remnant flux transport.  Figure~\ref{F-TL}b is compiled of zonally averaged magnetic fields which are chronologically ordered rotation by rotation. Positive/negative polarities are shown in blue/red. The colorbar saturates at $\pm$±1.0 G. Blue/red contours indicate domains of magnetic fields which exceed 0.5 G in absolute value. The time-latitude diagram demonstrates cyclic variations in zonal structure of solar magnetic fields and characterizes also their long-term behavior.  During  epochs of activity maxima, global rearrangements of the magnetic fields occur.

Most of the Sun's magnetic flux first appears as ARs which tend to concentrate within activity complexes (ACs) \citep{Gaizauskas83}. Long-lived ACs represent persistent patterns in magnetic-flux emergence \citep{Schrijver00}. The zones of intense sunspot activity are shown as black spots which are overlapped on the time-latitude distribution of magnetic fields.  They are identified using zonally averaged sunspot areas which exceed the threshold of  80  $\mu$hem/deg. As the cycle progresses, zones of sunspot activity migrate toward the equator according to Spoerer's law.
As the ACs evolved and after their decay, the remnant magnetic fields were dissipated and formed UMRs.
The diffusion and advection of the remnant flux determine the further evolution of  UMRs \citep{DeVore85, Wang91}. The UMRs of predominantly  following polarities are transported poleward due to meridional flows \citep{Hathaway11}.

This diagram shows cyclic rearrangements of large-scale magnetic fields and  polar-field reversals. After the decay of a large AC, its magnetic fields are dispersed into the surrounding photosphere. Long-lived ACs produce the remnant flux that form an extensive surge approaching the polar zones. The remnant flux surges are seen as inclined features which are stretched from mid- to high-latitudes (Figure~\ref{F-TL}b). They are shaped by  differential rotation and meridional flows. In each hemisphere,
several surges were usually formed  during an 11-yr activity cycle.
The approach of the most extensive surges to the Sun's poles reverses the polar-field. These events are marked by solid yellow arrows in Figure~\ref{F-TL}b. First surges are usually caused by ARs appearing at latitudes of about $\pm$30$^\circ$. The shape of these arrows typically corresponds to the averaged velocity profile of meridional flows \citep{Hathaway11}.

Large domains of highly concentrated magnetic flux show long-living ACs in Cycles  21--23.  Subsequent surges of following polarities strengthened the high-latitude magnetic fields of new polarities and widened their latitudinal extent. These surges are also marked with solid arrows (Figure~\ref{F-TL}b). The correspondent magnetic flux changes in $B_{\pm75}$ are marked with  vertical solid arrows in Figures~\ref{F-TL}a,c.

According to the Babcock-Leighton concept, the leading-polarity UMRs migrate to the equator where the opposite polarities are annihilated due to their reconnection. In the time-latitude diagram, however, the leading-polarity surges are sometimes transported polewards in Cycles 21--24. They are marked by dashed yellow arrows in Figure~\ref{F-TL}b. The poleward transport of opposite magnetic fields results in their cancellation and leads to a decrease in total polar flux \citep{Kitchatinov18}.

The extensive following-polarity surges seen in (Figure~\ref{F-TL}b) accelerate changes in the polar-fields (solid arrows, Figures~\ref{F-TL}a,c). The leading-polarity surges result in local maxima or minima in the polar-field buildup according to their polarity (dashed arrows, Figures~\ref{F-TL}a,c). As a rule, the leading-polarity surges originated after the decay of ARs which are characterized by a high magnetic flux density and non-Joy's tilts \citep{Yeates15,  Mordvinov15}. At high latitudes, leading-polarity surges migrate faster than one should expect in accordance with the typical velocity of meridional flows \citep{Hathaway11}. Such a fast propagation of open and closed magnetic fields is possible because of their interchange reconnection \citep{Wang04}.

In this study, we analyzed sunspot group tilts $\gamma_i$, $i$=1, 2, $\ldots$, 84226 taken from the DPD catalog \citep{Gyori04, Baranyi15} for 1974--2017. We considered the sunspot group tilts in a time-latitude aspect.
The scattered distribution of the tilt angles was averaged at each point over its vicinity, which covers 0.87 years in time and 2$^\circ$ in latitude.
 We treat ARs as test magnets and average their tilts without of any weights.
The local means or medians of noisy and scattered data provide robust estimates of the averaged tilt distribution. The domains of negative tilt predominance are shown in green (Figure~\ref{F-TL}b).

 Anti-Hale's ARs are shown by yellow markers.
They were identified using the catalog of bipolar magnetic regions for Cycle 21 \citep{Sheeley16}.
In Cycles 22, 23, anti-Hale's ARs were identified using Mount Wilson Observatory sunspot records \citep{Howard91}
and daily magnetograms which were derived by \cite{Sokoloff15}.
A similar technique was applied for {\it Helioseismic and Magnetic Imager}/{\it Solar Dynamic Observatory} data to determine anti-Hale's ARs in Cycle 24.

 During solar activity maxima, bipolar ARs of different orientation coexist at the same latitudes. However, our approach and the data used are not able to identify the effects of individual ARs. The time-latitude distributions show zonally averaged magnetic fields and AR tilts. We can identify a combined effect of multiple ARs. Generally, remnant flux surges are formed due to dispersal and interaction of mixed magnetic fields and the further poleward transport of dominant magnetic polarities.

 The DPD catalog includes about 75500 ARs which obey Joy's law. After their decay, UMRs of following polarities are usually formed at higher latitudes. The remnant flux of these ARs contributes most of the polar-flux buildup. Regardless of their east-west orientation, anti-Hale's ARs of appropriate north-south orientation contribute a minor remnant flux of the same polarity. The decay of non-Joy's and anti-Hale's ARs of the same north-south orientation results in leading-polarity remnant flux. The decay of anti-Hale's ARs of opposite north-south orientation led to the remnant flux weakening due to cancellation of opposite magnetic polarities.

 Taking into account the variety of AR configurations, it is reasonable to describe cyclic patterns in solar activity using universal terms. We studied dominant polarities in large-scale magnetic patterns in terms of Hale's law. The term ``following-polarity'' surge indicates a predominance of ARs which obey Hale's and Joy's laws while the term ``leading-polarity'' surge indicates a predominance of abnormal ARs and possible deviations from the Babcock-Leighton scenario. Surges of following- and leading-polarities correspond to the regular and stochastic components in remnant flux that is transported to the Sun's polar zones \citep{Kitchatinov18}.

In Cycle 21 multiple reversals of the Sun's polar fields were observed in both hemispheres.
At the North Pole, the polar-field changed its dominant polarity from positive to negative by early-1980.
Then, the leading-polarity surge approached the pole by early-1981 and resulted in short-term predominance of positive (leading) polarity (Figure~\ref{F-TL}b). The poleward transport of positive polarity led to the corresponding increase in $B_{75}$ shown with dashed arrow in Figure~\ref{F-TL}a. The leading-polarity surge originated due to the decay of the long-lived ACs which occurred at latitudes 10--15$^\circ$ throughout 1979--1980 (Figure~\ref{F-TL}b). These ACs were characterized by the well-defined domains of non-Joy's tilt.

The main domain was surrounded by anti-Hale's ARs (marked yellow). Positive (following) polarity UMRs originated from the decay of non-Joy's ARs.
Positive polarity also resulted from the decay of anti-Hale's ARs which surround the domain of negative tilt. In this particular case, the decay of anti-Hale's ARs and non-Joy's ARs results in a positive polarity surge that reached the North Pole and led to the second reversal (Figures~\ref{F-TL}b, ~\ref{F-C21}b).
This surge was marked with a dashed arrow taking into account its starting point.
The next surge of following (negative) polarity changed the dominant polarity from positive to negative at the North Pole.

At the South Pole multiple reversals occurred in late-1980, early-1982 and late-1982. The leading-polarity surge originated due to the decay of ACs with abnormal tilts which concentrated at 5--30$^\circ$ for 1981--1982 (Figure~\ref{F-TL}b). The poleward transport of negative-polarity flux changed the dominant polarity from positive to negative in 1982. The incoming flux of opposite polarity is also visible in $B_{-75}$ as the local minimum marked with the dashed arrow in Figure~\ref{F-TL}c.

Surges of opposite magnetic polarities were also observed in both hemispheres throughout Cycle 22 (Figure~\ref{F-TL}b).
Unlike Cycle 21, the domains of negative tilt predominance were less concentrated in the southern hemisphere.
Under such conditions, a three-fold reversal of the polar fields was hardly noticeable in the southern hemisphere.
The leading polarities led to a disappearance of polar magnetic flux at both poles, however, the incoming fluxes were insufficient to change the dominant magnetic polarities.

Leading-polarity surges occurred in Cycles 23--24 also (dashed arrows, Figure~\ref{F-TL}b). Their effects on high-latitude fields are also marked with dashed arrows (Figures~\ref{F-TL}a,c). Their possible sources were identified using high-resolution synoptic maps \citep{Mordvinov16, Lockwood17, Kitchatinov18}.

Cycle 24 was characterized by a low activity that evolved asynchronously in the northern and southern hemispheres. Emergent magnetic flux peaked in late 2011 in the northern hemisphere.
The decaying ARs resulted in dominant flux of following (positive) polarity that leads to polar-field reversal by early 2013 \citep{Mordvinov14, Sun15, Petrie17}.
The crucial surge is marked with the solid arrow, while two adjacent surges of leading (negative) polarity  are marked with dashed arrows.
Prolonged surges of positive polarity originated due to the decay of small ARs that existed in the northern hemisphere during 2014--2015.

Thus, the time-latitude diagram depicts long-living patterns of emergent magnetic flux and its decay that led to the formation of remnant-flux surges.
The poleward transport of UMRs of mixed magnetic polarities, their further cancellation and accumulation led to the polar-field buildup.
Global rearrangements of solar magnetic fields and  polar-field reversals make evident the causal relations between the local and large-scale magnetic fields. The time-latitude distribution of emergent magnetic flux and its further transformations determine the evolution of the Sun's polar fields and individual properties of Cycles 21--24.

\section{A Close Look at the Polar Field Reversals in Cycle 21}
\label{S-Cyc21}

\begin{figure}    %%%%%%%%%%%%%%%%%% FIGURE 2
\centerline{\includegraphics[width=0.9\textwidth,clip=]{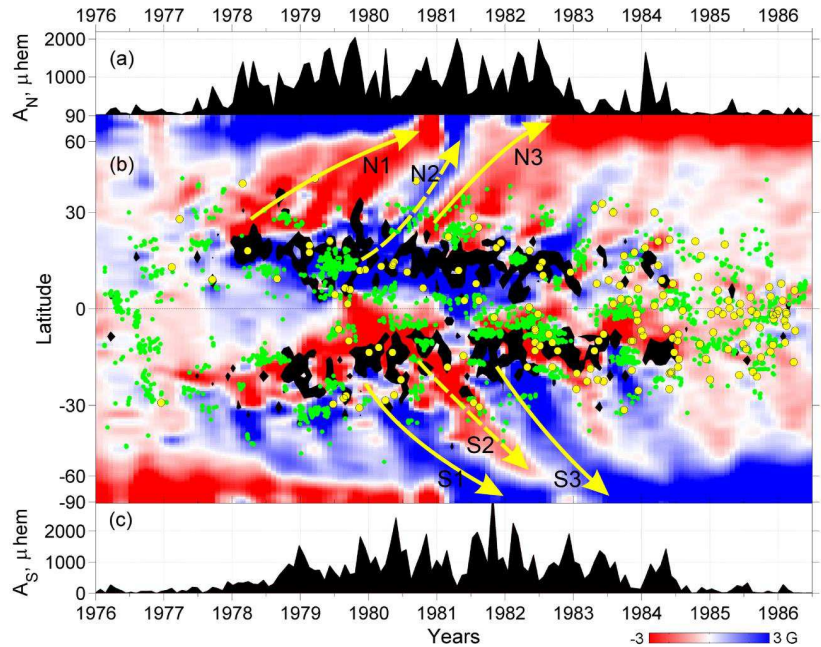}}
  \caption{Changes in sunspot areas in the northern (a) and southern (c) hemispheres for 1976--1986. The time-latitude distribution of the averaged magnetic fields is shown in red-to-blue (b). Zones of intense sunspot activity,  domains of negative tilt and anti-Hale's ARs are shown in black, green, and yellow, respectively.}

   \label{F-C21}
   \end{figure}

To study the polar-field reversals in Cycle 21 we analyzed synoptic maps from Kitt Peak Observatory \citep{Jones92}. The synoptic maps of the photospheric magnetic fields are uniformly sampled over the longitude and sine latitude (360$\times$180 pixels). These maps present measurements of solar magnetic fields at higher resolution on an independent scale
\url{ftp://vso.nso.edu/kpvt/synoptic/mag}. Figures~\ref{F-C21}a,c show changes in sunspot areas in the northern (a) and southern (c) hemispheres. Figure~\ref{F-C21}b depicts zonally averaged magnetic fields in red-to-blue by analogy with Figure~\ref{F-TL}b.

 The time-latitude diagram shows the evolution of large-scale magnetic fields in more detail. The crucial surges N1/S1 resulted in the polar-field reversals at the northern/southern poles. Zones of intense sunspot activity are shown in black, they correspond to sunspot area density $>$ 100 $\mu$hem/deg. The leading-polarity surges N2/S2 led to the short-term predominance of positive/negative polarities at the northern/southern poles. Subsequent surges N3 and S3 changed the signs of the polar-fields as these were after the first reversal.  The domains of negative tilt predominance are shown in green.
 Anti-Hale's ARs are marked with yellow markers.
 The degree of tilt averaging corresponds to a spatio-temporal element of size 0.26 years in time and 2$^\circ$ in latitude.

The time-latitude diagram (Figure~\ref{F-C21}b) depicts the evolution of the large-scale magnetic fields in more detail. The distribution of emergent flux, its further decay and formation of the remnant flux surges agree with spatio-temporal patterns seen in Figure~\ref{F-TL}b. The crucial surges of  following and leading polarities resulted in peculiar features of Cycle 21 and  multiple polar-field reversal at the northern and southern poles.
It of interest that ARs with ``incorrect'' orientations contribute 37\% of total magnetic flux erupted during Cycle 21 \citep{Wang89}.

It should be noted that previous analysis of the poleward migration of chromospheric filaments \citep{Makarov83} revealed no  triple polar field reversals in Cycle 21.
Possibly, stable filament do not have time to form under fast changes in dominant magnetic polarities.

\section{A Triple Polar Field Reversal in Cycle 20}
\label{S5-Cycle20}

The Sun's magnetic activity varies on a secular timescale. During the first half of the 20th century, amplitudes of the 11-year cycles tended to increase.  After the highest Cycle 19 a low-amplitude Cycle 20 occurred for 1964--1976. Since then, the level of magnetic activity has decreased.
This long-term variability comprises a complete secular (Gleissberg) cycle. Taking into account the abrupt change in the long-term tendency, it is of interest to study evolution of magnetic activity after Cycle 19.

The effects of magnetic field structure are also observed in the Sun's chromosphere. Large unipolar magnetic fields are outlined by chromospheric filaments which trace the lines of polarity inversions. Long-term chromospheric observations were compiled in the synoptic maps of large-scale magnetic fields \citep{McIntosh79}. \cite{Makarov83} analyzed similar synoptic maps for 1945--1981. They studied the global evolution of magnetic fields and found three-fold reversals of the polar field in Cycles 19, 20 in the northern hemisphere. So far, there is no convincing physical explanation for these phenomena.

 \begin{figure}    %%%%%%%%%%%%%%%%%% FIGURE 3
\centerline{\includegraphics[width=0.9\textwidth,clip=]{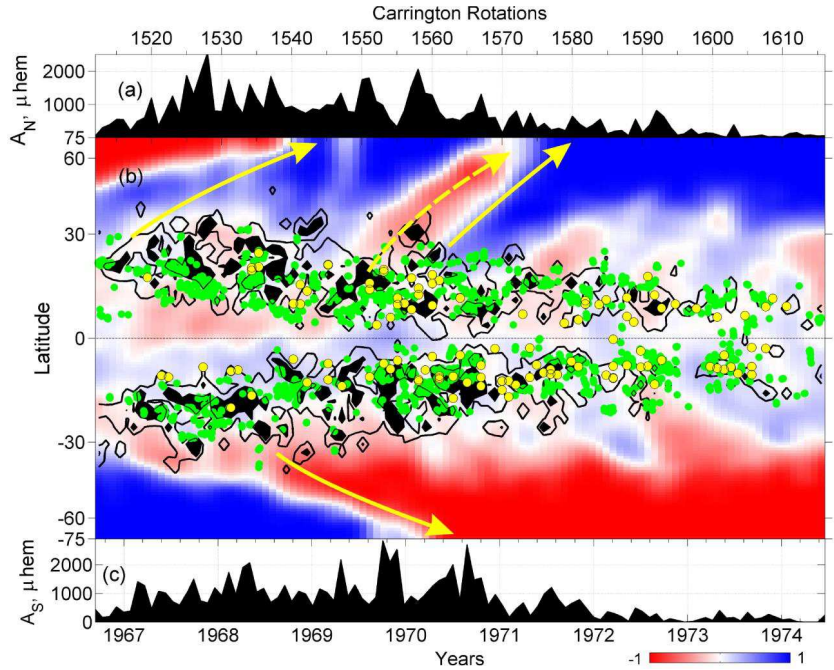}
              }
  \caption{Changes in sunspot areas in  the northern (a) and southern (c) hemispheres for 1966--1974. The distribution of zonally averaged values sign($B$) depicts time-latitude behavior of large-scale magnetic fields in red-to-blue (b). Zones of intense sunspot activity, domains of negative tilt, and anti-Hale's ARs are superimposed in black, green, and yellow, respectively.
  The solid/dashed  arrows depict the remnant-flux surges of following/leading polarities.}

   \label{F-C20}
   \end{figure}

In Cycle 20, sunspot activity developed asynchronously in the northern and southern hemispheres. The sunspot areas peaked in 1967/1969 in the northern/southern hemispheres (Figures~\ref{F-C20}a,c).
We analyzed synoptic maps from the \cite{McIntosh79} archive to study the time-latitude distribution of the Sun's large-scale magnetic fields in Cycle 20. The proxy maps satisfactorily agree with synoptic magnetograms \citep{Snodgrass00}. The McIntosh archive covers the period 1964–-2009 \citep{Webb17}.
%(https://www2.hao.ucar.edu/mcintosh-archive/four-cycles-solar-synoptic-maps).

 The H$_\alpha$ synoptic charts display lines of magnetic polarity inversion  which outline polarity boundaries of large-scale magnetic fields.
Polarities of magnetic patterns were inferred from magnetograms  \citep{McIntosh79, Makarov83, Makarov86}.
The digitized maps quantify magnetic patterns by $\pm$1 according to their polarities.
These maps show the distribution of  polarity signs  without regard to magnetic field strength.
The time-latitude distribution of sign($B$) value shows the appearance of dominant magnetic polarities taking into account their signs.

 Figure~\ref{F-C20}b shows a time-latitude diagram that is compiled of zonally averaged proxy maps for 1966--1974. The averaged values sign($B$) are shown in red-to-blue. The diagram depicts dominant magnetic polarities and their changes in Cycle 20. Zones of intense sunspot activity are shown in black, by analogy with Figure~\ref{F-TL}b. For the period 1966--1974, we use sunspot group tilt data from Mount Wilson Observatory \citep{Howard91}. Domains of negative tilt predominance were identified from the averaged tilt distribution. The degree of tilt averaging corresponds to a spatio-temporal element of size 0.26 years in time and 2$^\circ$ in latitude.

Both sunspot activity and emergent flux are characterized by significant North-South asymmetry. At the South Pole, the polar-field reversed at about 1970.2 due to the remnant-flux surge of negative polarity marked with a solid arrow (Figure ~\ref{F-C20}b). At the North Pole, the polar-field reversed first at 1968.8 changing its sign from negative to positive. It was  a regular reversal due to the poleward transport the following (positive) polarity.

In the northern hemisphere, multiple long-lived ACs occurred in 1967--1970. A huge surge of leading polarity was formed after the decay of the large activity centers in 1969--1970.
 The domains of non-Joy's tilt surrounded the activity center in 1969.
 Anti-Hale's ARs also occurred within the low-latitude base of the negative-polarity surge.
It is under such conditions that surges of leading polarity are usually formed. The remnant flux of negative polarity was transported poleward and resulted in the second reversal. Negative polarity dominated at the North Pole during 1971.0--1971.5.

 The decay of subsequent ACs resulted in the next surge of following (positive) polarity that led to the third reversal  at the North Pole.
In the southern hemisphere, the decay of usual ARs produced an extensive surge of negative polarity that reached the South Pole and led to a regular reversal.
These changes in dominant magnetic polarities at the Sun's  poles  agree with the results derived from the poleward migration of chromospheric filaments in Cycle 20 \citep{Makarov83}.

\section{Conclusions}
\label{S6-Conc}
Synoptic magnetograms and their analogs from the McIntosh archive were analyzed to study the Sun's polar-field buildup in Cycles 20--24. Evolution of the polar magnetic field of the Sun was studied in relation to cyclic changes in its sunspot activity and large-scale magnetic fields.
The extended analysis improved our understanding of polar field evolution with new essential detail. Thus, the causal and evolutionary relations between  decaying ACs, remnant flux surges and polar magnetic field buildup are evident in their time-latitude behavior.

Peculiar properties of recent cycles have shown that the polar field buildup depends on emergent magnetic flux, its further dispersal and  poleward transport. These processes determine inherent time-latitude patterns which vary from cycle to cycle. Following-polarity surges originate after the decay of long-lived activity centers where ARs with positive (Joy's) tilts dominate. Crucial surges of  following polarity resulted in the polar-field reversals. Subsequent surges of following-polarities led to a strengthening of the polar flux.

Leading-polarity surges originate with the decay of ARs which violate Joy's and Hale's laws.
In averaged time-latitude distributions of tilt angles, multiple domains of negative tilt predominance were found.
 Persistent domains of negative tilt predominance are possibly caused by some anomalies in the subphotospheric toroidal field which appear also in recurrent magnetic bipoles in the upper atmosphere.
Sometimes anti-Hale's ARs concentrate near or within domains of negative tilt predominance.
Generally, it is difficult to distinguish the effect of  non-Joy's and anti-Hale's ARs.
 A combined effect of non-Joy's and anti-Hale's ARs resulted remnant-flux surges of leading polarities (in terms of Hale's law).
  The poleward transport of opposite magnetic polarities led to the multiple reversals in Cycles 20 and 21.
The cancellation of opposite magnetic polarities  results in the Sun's polar flux  weakening in Cycles 21--24.

%%%%%%%%%%%%%%%%%%%%%%%%%%%%%%%%%%%%%%%%%%%%%%%%%%%%%%%%%%%%%%%%%%%%%%%%%%%
\begin{acks}

This research utilizes synoptic magnetograms from WSO and Kitt Peak observatories. We used also sunspot group tilt data from Mount Wilson and Debrecen observatories.
The synoptic maps from the McIntosh Archive were also used in this research.
 We are grateful to A.I. Khlystova for preparing data on anti-Hale's active regions, and J.~Sutton for improving the English version of the manuscript.
The work was supported by Basic Research program II.16 and the RFBR  project 17-02-00016.

\end{acks}

\section{Disclosure of Potential Conflicts of Interest}
The authors declare that they have no conflicts of interest.

%\noindent To change a title use an optional parameter:\par
%\verb+\begin{acks}[Acknowledgements]...\end{acks}+

%\acknowledgment US spelling: \verb+\acknowledgment+
%\acknowledgement British  spelling: \verb+\acknowledgement+

%%%%%%%%%%%%%%%%%%%%%%%%%%%%%%%%%%%%%%%%%%%%%%%%%%%%%%%%%%%%%%%%%%%%%%%%%%%
     % format of references provided by the journal (.bst)
\bibliographystyle{spr-mp-sola}

\begin{thebibliography}{57}
% BibTex style file: spr-mp-sola.bst (nameyear), 2018-10-11
\ifx\bisbn     \undefined \def\bisbn  #1{ISBN #1}\fi
\ifx\binits    \undefined \def\binits#1{#1}\fi
\ifx\bauthor   \undefined \def\bauthor#1{#1}\fi
\ifx\batitle   \undefined \def\batitle#1{#1}\fi
\ifx\bjtitle   \undefined \def\bjtitle#1{\textit{#1}}\fi
\ifx\bvolume   \undefined \def\bvolume#1{\textbf{#1}}\fi
\ifx\byear     \undefined \def\byear#1{#1}\fi
\ifx\bissue    \undefined \def\bissue#1{#1}\fi
\ifx\bfpage    \undefined \def\bfpage#1{#1}\fi
\ifx\blpage    \undefined \def\blpage #1{#1}\fi
\ifx\burl      \undefined \def\burl#1{\textsf{#1}}\fi
\ifx\href      \undefined \def\href#1#2{\textsf{#2}}\fi
\ifx\betal     \undefined \def\betal{\textit{et al.}}\fi
\ifx\bctitle   \undefined \def\bctitle#1{#1}\fi
\ifx\beditor   \undefined \def\beditor#1{#1}\fi
\ifx\bbtitle   \undefined \def\bbtitle#1{\textit{#1}}\fi
\ifx\bedition  \undefined \def\bedition#1{#1}\fi
\ifx\bseriesno \undefined \def\bseriesno#1{\textbf{#1}}\fi
\ifx\blocation \undefined \def\blocation#1{#1}\fi
\ifx\bsertitle \undefined \def\bsertitle#1{\textit{#1}}\fi
\ifx\bsnm      \undefined \def\bsnm#1{#1}\fi
\ifx\bsuffix   \undefined \def\bsuffix#1{#1}\fi
\ifx\bparticle \undefined \def\bparticle#1{#1}\fi
\ifx\barticle  \undefined \def\barticle#1{}\fi
\ifx\binstitute  \undefined \def\binstitute#1{#1}\fi
\ifx\bpublisher  \undefined \def\bpublisher#1{#1}\fi
\ifx\doiurl    \undefined
  \def\doiurl#1{\href{https://doi.org/#1}{\textsf{DOI}}}\fi
\ifx\arxivurl  \undefined
  \def\arxivurl#1{\href{http://arxiv.org/abs/#1}{\textsf{arXiv}}}\fi
\ifx\adsurl    \undefined
  \def\adsurl#1{\href{http://adsabs.harvard.edu/abs/#1}{\textsf{ADS}}}\fi
\ifx\botherref \undefined \def\botherref#1{}\fi
\ifx\url       \undefined \def\url#1{\textsf{#1}}\fi
\ifx\bchapter  \undefined \def\bchapter#1{}\fi
\ifx\bbook     \undefined \def\bbook#1{}\fi
\ifx\bcomment  \undefined \def\bcomment#1{#1}\fi
\ifx\oauthor   \undefined \def\oauthor#1{#1}\fi
\ifx\citeauthoryear \undefined\def \citeauthoryear#1{#1}\fi
\def\endbibitem {}
\ifx\bconflocation  \undefined \def\bconflocation#1{#1} \fi

\bibitem[\protect\citeauthoryear{{Babcock}}{1961}]{Babcock61}
\begin{barticle}
\bauthor{\bsnm{{Babcock}}, \binits{H.W.}}:
\byear{1961},
\batitle{{The Topology of the Sun's Magnetic Field and the 22-YEAR Cycle.}}
\bjtitle{\apj}
\bvolume{133},
\bfpage{572}.
\doiurl{https://doi.org/10.1086/147060}.
\adsurl{1961ApJ...133..572B}.
\end{barticle}
\endbibitem

\bibitem[\protect\citeauthoryear{{Baranyi}}{2015}]{Baranyi15}
\begin{barticle}
\bauthor{\bsnm{{Baranyi}}, \binits{T.}}:
\byear{2015},
\batitle{{Comparison of Debrecen and Mount Wilson/Kodaikanal sunspot group tilt
  angles and the Joy's law}}.
\bjtitle{\mnras}
\bvolume{447},
\bfpage{1857}.
\doiurl{https://doi.org/10.1093/mnras/stu2572}.
\adsurl{2015MNRAS.447.1857B}.
\end{barticle}
\endbibitem

\bibitem[\protect\citeauthoryear{{Baumann} \textit{et~al.}}{2004}]{Baumann04}
\begin{barticle}
\bauthor{\bsnm{{Baumann}}, \binits{I.}},
\bauthor{\bsnm{{Schmitt}}, \binits{D.}},
\bauthor{\bsnm{{Sch{\"u}ssler}}, \binits{M.}},
\bauthor{\bsnm{{Solanki}}, \binits{S.K.}}:
\byear{2004},
\batitle{{Evolution of the large-scale magnetic field on the solar surface: A
  parameter study}}.
\bjtitle{\aap}
\bvolume{426},
\bfpage{1075}.
\doiurl{https://doi.org/10.1051/0004-6361:20048024}.
\adsurl{2004A26A...426.1075B}.
\end{barticle}
\endbibitem

\bibitem[\protect\citeauthoryear{{Cameron} \textit{et~al.}}{2018}]{Cameron18}
\begin{barticle}
\bauthor{\bsnm{{Cameron}}, \binits{R.H.}},
\bauthor{\bsnm{{Duvall}}, \binits{T.L.}},
\bauthor{\bsnm{{Sch{\"u}ssler}}, \binits{M.}},
\bauthor{\bsnm{{Schunker}}, \binits{H.}}:
\byear{2018},
\batitle{{Observing and modeling the poloidal and toroidal fields of the solar
  dynamo}}.
\bjtitle{\aap}
\bvolume{609},
\bfpage{A56}.
\doiurl{https://doi.org/10.1051/0004-6361/201731481}.
\adsurl{2018A26A...609A..56C}.
\end{barticle}
\endbibitem

\bibitem[\protect\citeauthoryear{{Choudhuri} and {Karak}}{2012}]{Choudhuri12}
\begin{barticle}
\bauthor{\bsnm{{Choudhuri}}, \binits{A.R.}},
\bauthor{\bsnm{{Karak}}, \binits{B.B.}}:
\byear{2012},
\batitle{{Origin of Grand Minima in Sunspot Cycles}}.
\bjtitle{Physical Review Letters}
\bvolume{109}(\bissue{17}),
\bfpage{171103}.
\doiurl{https://doi.org/10.1103/PhysRevLett.109.171103}.
\adsurl{2012PhRvL.109q1103C}.
\end{barticle}
\endbibitem

\bibitem[\protect\citeauthoryear{{Choudhuri}, {Schussler}, and
  {Dikpati}}{1995}]{Choudhuri95}
\begin{barticle}
\bauthor{\bsnm{{Choudhuri}}, \binits{A.R.}},
\bauthor{\bsnm{{Schussler}}, \binits{M.}},
\bauthor{\bsnm{{Dikpati}}, \binits{M.}}:
\byear{1995},
\batitle{{The solar dynamo with meridional circulation.}}
\bjtitle{\aap}
\bvolume{303},
\bfpage{L29}.
\adsurl{1995A26A...303L..29C}.
\end{barticle}
\endbibitem

\bibitem[\protect\citeauthoryear{{Dasi-Espuig} \textit{et~al.}}{2010}]{Dasi10}
\begin{barticle}
\bauthor{\bsnm{{Dasi-Espuig}}, \binits{M.}},
\bauthor{\bsnm{{Solanki}}, \binits{S.K.}},
\bauthor{\bsnm{{Krivova}}, \binits{N.A.}},
\bauthor{\bsnm{{Cameron}}, \binits{R.}},
\bauthor{\bsnm{{Pe{\~n}uela}}, \binits{T.}}:
\byear{2010},
\batitle{{Sunspot group tilt angles and the strength of the solar cycle}}.
\bjtitle{\aap}
\bvolume{518},
\bfpage{A7}.
\doiurl{https://doi.org/10.1051/0004-6361/201014301}.
\adsurl{2010A26A...518A...7D}.
\end{barticle}
\endbibitem

\bibitem[\protect\citeauthoryear{{Dasi-Espuig} \textit{et~al.}}{2013}]{Dasi13}
\begin{barticle}
\bauthor{\bsnm{{Dasi-Espuig}}, \binits{M.}},
\bauthor{\bsnm{{Solanki}}, \binits{S.K.}},
\bauthor{\bsnm{{Krivova}}, \binits{N.A.}},
\bauthor{\bsnm{{Cameron}}, \binits{R.}},
\bauthor{\bsnm{{Pe{\~n}uela}}, \binits{T.}}:
\byear{2013},
\batitle{{Sunspot group tilt angles and the strength of the solar cycle
  (Corrigendum)}}.
\bjtitle{\aap}
\bvolume{556},
\bfpage{C3}.
\doiurl{https://doi.org/10.1051/0004-6361/201014301e}.
\adsurl{2013A26A...556C...3D}.
\end{barticle}
\endbibitem

\bibitem[\protect\citeauthoryear{{DeVore} \textit{et~al.}}{1985}]{DeVore85}
\begin{barticle}
\bauthor{\bsnm{{DeVore}}, \binits{C.R.}},
\bauthor{\bsnm{{Sheeley}}, \binits{N.R.} \bsuffix{Jr.}},
\bauthor{\bsnm{{Boris}}, \binits{J.P.}},
\bauthor{\bsnm{{Young}}, \binits{T.R.} \bsuffix{Jr.}},
\bauthor{\bsnm{{Harvey}}, \binits{K.L.}}:
\byear{1985},
\batitle{{Simulations of magnetic-flux transport in solar active regions}}.
\bjtitle{\solphys}
\bvolume{102},
\bfpage{41}.
\doiurl{https://doi.org/10.1007/BF00154036}.
\adsurl{1985SoPh..102...41D}.
\end{barticle}
\endbibitem

\bibitem[\protect\citeauthoryear{{Durney}}{1995}]{Durney95}
\begin{barticle}
\bauthor{\bsnm{{Durney}}, \binits{B.R.}}:
\byear{1995},
\batitle{{On a Babcock-Leighton dynamo model with a deep-seated generating
  layer for the toroidal magnetic field}}.
\bjtitle{\solphys}
\bvolume{160},
\bfpage{213}.
\doiurl{https://doi.org/10.1007/BF00732805}.
\adsurl{1995SoPh..160..213D}.
\end{barticle}
\endbibitem

\bibitem[\protect\citeauthoryear{{Duvall} \textit{et~al.}}{1977}]{Duvall77}
\begin{botherref}
\oauthor{\bsnm{{Duvall}}, \binits{T.L.} \bsuffix{Jr.}},
\oauthor{\bsnm{{Wilcox}}, \binits{J.M.}},
\oauthor{\bsnm{{Svalgaard}}, \binits{L.}},
\oauthor{\bsnm{{Scherrer}}, \binits{P.H.}},
\oauthor{\bsnm{{McIntosh}}, \binits{P.S.}}:
1977,
{Comparison of H alpha synoptic charts with the large-scale solar magnetic
  field as observed at Stanford}.
\textit{NASA STI/Recon Technical Report N}
\textbf{77}.
\adsurl{1977STIN...7729049D}.
\end{botherref}
\endbibitem

\bibitem[\protect\citeauthoryear{{Erofeev}}{2004}]{Erofeev04}
\begin{bchapter}
\bauthor{\bsnm{{Erofeev}}, \binits{D.V.}}:
\byear{2004},
\bctitle{{An observational evidence for the Babcock-Leighton dynamo scenario}}.
In: \beditor{\bsnm{{Stepanov}}, \binits{A.V.}},
\beditor{\bsnm{{Benevolenskaya}}, \binits{E.E.}},
\beditor{\bsnm{{Kosovichev}}, \binits{A.G.}} (eds.)
\bbtitle{Multi-Wavelength Investigations of Solar Activity},
\bsertitle{IAU Symposium}
\bseriesno{223},
\bfpage{97}.
\doiurl{https://doi.org/10.1017/S1743921304005228}.
\adsurl{2004IAUS..223...97E}.
\end{bchapter}
\endbibitem

\bibitem[\protect\citeauthoryear{{Gaizauskas}
  \textit{et~al.}}{1983}]{Gaizauskas83}
\begin{barticle}
\bauthor{\bsnm{{Gaizauskas}}, \binits{V.}},
\bauthor{\bsnm{{Harvey}}, \binits{K.L.}},
\bauthor{\bsnm{{Harvey}}, \binits{J.W.}},
\bauthor{\bsnm{{Zwaan}}, \binits{C.}}:
\byear{1983},
\batitle{{Large-scale patterns formed by solar active regions during the
  ascending phase of cycle 21}}.
\bjtitle{\apj}
\bvolume{265},
\bfpage{1056}.
\doiurl{https://doi.org/10.1086/160747}.
\adsurl{1983ApJ...265.1056G}.
\end{barticle}
\endbibitem

\bibitem[\protect\citeauthoryear{{Golubeva} and {Mordvinov}}{2016}]{Golubeva16}
\begin{barticle}
\bauthor{\bsnm{{Golubeva}}, \binits{E.M.}},
\bauthor{\bsnm{{Mordvinov}}, \binits{A.V.}}:
\byear{2016},
\batitle{{Decay of Activity Complexes, Formation of Unipolar Magnetic Regions,
  and Coronal Holes in Their Causal Relation}}.
\bjtitle{\solphys}
\bvolume{291},
\bfpage{3605}.
\doiurl{https://doi.org/10.1007/s11207-016-1011-1}.
\adsurl{2016SoPh..291.3605G}.
\end{barticle}
\endbibitem

\bibitem[\protect\citeauthoryear{{Golubeva} and {Mordvinov}}{2017}]{Golubeva17}
\begin{barticle}
\bauthor{\bsnm{{Golubeva}}, \binits{E.M.}},
\bauthor{\bsnm{{Mordvinov}}, \binits{A.V.}}:
\byear{2017},
\batitle{{Rearrangements of Open Magnetic Flux and Formation of Polar Coronal
  Holes in Cycle 24}}.
\bjtitle{\solphys}
\bvolume{292},
\bfpage{175}.
\doiurl{https://doi.org/10.1007/s11207-017-1200-6}.
\adsurl{2017SoPh..292..175G}.
\end{barticle}
\endbibitem

\bibitem[\protect\citeauthoryear{{Gy{\"o}ri} \textit{et~al.}}{2004}]{Gyori04}
\begin{barticle}
\bauthor{\bsnm{{Gy{\"o}ri}}, \binits{L.}},
\bauthor{\bsnm{{Baranyi}}, \binits{T.}},
\bauthor{\bsnm{{Ludm{\'a}ny}}, \binits{A.}},
\bauthor{\bsnm{{Gerlei}}, \binits{O.}},
\bauthor{\bsnm{{Csepura}}, \binits{G.}},
\bauthor{\bsnm{{Mez{\"o}}}, \binits{G.}}:
\byear{2004},
\batitle{{Debrecen Photoheliographic Data for the Years 1993-1995}}.
\bjtitle{Publications of Debrecen Heliophysical Observatory}
\bvolume{17},
\bfpage{1}.
\adsurl{2004PDHO...17....1G}.
\end{barticle}
\endbibitem

\bibitem[\protect\citeauthoryear{{Hathaway} and {Rightmire}}{2011}]{Hathaway11}
\begin{barticle}
\bauthor{\bsnm{{Hathaway}}, \binits{D.H.}},
\bauthor{\bsnm{{Rightmire}}, \binits{L.}}:
\byear{2011},
\batitle{{Variations in the Axisymmetric Transport of Magnetic Elements on the
  Sun: 1996-2010}}.
\bjtitle{\apj}
\bvolume{729},
\bfpage{80}.
\doiurl{https://doi.org/10.1088/0004-637X/729/2/80}.
\adsurl{2011ApJ...729...80H}.
\end{barticle}
\endbibitem

\bibitem[\protect\citeauthoryear{{Hoeksema}}{2010}]{Hoeksema10}
\begin{bchapter}
\bauthor{\bsnm{{Hoeksema}}, \binits{J.T.}}:
\byear{2010},
\bctitle{{Evolution of the large-scale magnetic field over three solar
  cycles}}.
In: \beditor{\bsnm{{Kosovichev}}, \binits{A.G.}},
\beditor{\bsnm{{Andrei}}, \binits{A.H.}},
\beditor{\bsnm{{Rozelot}}, \binits{J.-P.}} (eds.)
\bbtitle{Solar and Stellar Variability: Impact on Earth and Planets},
\bsertitle{IAU Symposium}
\bseriesno{264},
\bfpage{222}.
\doiurl{https://doi.org/10.1017/S1743921309992675}.
\adsurl{2010IAUS..264..222H}.
\end{bchapter}
\endbibitem

\bibitem[\protect\citeauthoryear{{Howard}}{1991}]{Howard91}
\begin{barticle}
\bauthor{\bsnm{{Howard}}, \binits{R.F.}}:
\byear{1991},
\batitle{{Axial tilt angles of sunspot groups}}.
\bjtitle{\solphys}
\bvolume{136},
\bfpage{251}.
\doiurl{https://doi.org/10.1007/BF00146534}.
\adsurl{1991SoPh..136..251H}.
\end{barticle}
\endbibitem

\bibitem[\protect\citeauthoryear{{Jiang}, {Cameron}, and
  {Sch{\"u}ssler}}{2015}]{Jiang15}
\begin{barticle}
\bauthor{\bsnm{{Jiang}}, \binits{J.}},
\bauthor{\bsnm{{Cameron}}, \binits{R.H.}},
\bauthor{\bsnm{{Sch{\"u}ssler}}, \binits{M.}}:
\byear{2015},
\batitle{{The Cause of the Weak Solar Cycle 24}}.
\bjtitle{\apjl}
\bvolume{808},
\bfpage{L28}.
\doiurl{https://doi.org/10.1088/2041-8205/808/1/L28}.
\adsurl{2015ApJ...808L..28J}.
\end{barticle}
\endbibitem

\bibitem[\protect\citeauthoryear{{Jiang} \textit{et~al.}}{2010}]{Jiang10}
\begin{barticle}
\bauthor{\bsnm{{Jiang}}, \binits{J.}},
\bauthor{\bsnm{{I{\c s}ik}}, \binits{E.}},
\bauthor{\bsnm{{Cameron}}, \binits{R.H.}},
\bauthor{\bsnm{{Schmitt}}, \binits{D.}},
\bauthor{\bsnm{{Sch{\"u}ssler}}, \binits{M.}}:
\byear{2010},
\batitle{{The Effect of Activity-related Meridional Flow Modulation on the
  Strength of the Solar Polar Magnetic Field}}.
\bjtitle{\apj}
\bvolume{717},
\bfpage{597}.
\doiurl{https://doi.org/10.1088/0004-637X/717/1/597}.
\adsurl{2010ApJ...717..597J}.
\end{barticle}
\endbibitem

\bibitem[\protect\citeauthoryear{{Jiang} \textit{et~al.}}{2013}]{Jiang13}
\begin{barticle}
\bauthor{\bsnm{{Jiang}}, \binits{J.}},
\bauthor{\bsnm{{Cameron}}, \binits{R.H.}},
\bauthor{\bsnm{{Schmitt}}, \binits{D.}},
\bauthor{\bsnm{{I{\c s}{\i}k}}, \binits{E.}}:
\byear{2013},
\batitle{{Modeling solar cycles 15 to 21 using a flux transport dynamo}}.
\bjtitle{\aap}
\bvolume{553},
\bfpage{A128}.
\doiurl{https://doi.org/10.1051/0004-6361/201321145}.
\adsurl{2013A26A...553A.128J}.
\end{barticle}
\endbibitem

\bibitem[\protect\citeauthoryear{{Jiang} \textit{et~al.}}{2014}]{Jiang14}
\begin{barticle}
\bauthor{\bsnm{{Jiang}}, \binits{J.}},
\bauthor{\bsnm{{Hathaway}}, \binits{D.H.}},
\bauthor{\bsnm{{Cameron}}, \binits{R.H.}},
\bauthor{\bsnm{{Solanki}}, \binits{S.K.}},
\bauthor{\bsnm{{Gizon}}, \binits{L.}},
\bauthor{\bsnm{{Upton}}, \binits{L.}}:
\byear{2014},
\batitle{{Magnetic Flux Transport at the Solar Surface}}.
\bjtitle{\ssr}
\bvolume{186},
\bfpage{491}.
\doiurl{https://doi.org/10.1007/s11214-014-0083-1}.
\adsurl{2014SSRv..186..491J}.
\end{barticle}
\endbibitem

\bibitem[\protect\citeauthoryear{{Jones} \textit{et~al.}}{1992}]{Jones92}
\begin{barticle}
\bauthor{\bsnm{{Jones}}, \binits{H.P.}},
\bauthor{\bsnm{{Duvall}}, \binits{T.L.} \bsuffix{Jr.}},
\bauthor{\bsnm{{Harvey}}, \binits{J.W.}},
\bauthor{\bsnm{{Mahaffey}}, \binits{C.T.}},
\bauthor{\bsnm{{Schwitters}}, \binits{J.D.}},
\bauthor{\bsnm{{Simmons}}, \binits{J.E.}}:
\byear{1992},
\batitle{{The NASA/NSO spectromagnetograph}}.
\bjtitle{\solphys}
\bvolume{139},
\bfpage{211}.
\doiurl{https://doi.org/10.1007/BF00159149}.
\adsurl{1992SoPh..139..211J}.
\end{barticle}
\endbibitem

\bibitem[\protect\citeauthoryear{{Kitchatinov}, {Mordvinov}, and
  {Nepomnyashchikh}}{2018}]{Kitchatinov18}
\begin{barticle}
\bauthor{\bsnm{{Kitchatinov}}, \binits{L.L.}},
\bauthor{\bsnm{{Mordvinov}}, \binits{A.V.}},
\bauthor{\bsnm{{Nepomnyashchikh}}, \binits{A.A.}}:
\byear{2018},
\batitle{{Modelling variability of solar activity cycles}}.
\bjtitle{\aap}
\bvolume{615},
\bfpage{A38}.
\doiurl{https://doi.org/10.1051/0004-6361/201732549}.
\adsurl{2018A26A...615A..38K}.
\end{barticle}
\endbibitem

\bibitem[\protect\citeauthoryear{{Kosovichev} and
  {Stenflo}}{2008}]{Kosovichev08}
\begin{barticle}
\bauthor{\bsnm{{Kosovichev}}, \binits{A.G.}},
\bauthor{\bsnm{{Stenflo}}, \binits{J.O.}}:
\byear{2008},
\batitle{{Tilt of Emerging Bipolar Magnetic Regions on the Sun}}.
\bjtitle{\apjl}
\bvolume{688},
\bfpage{L115}.
\doiurl{https://doi.org/10.1086/595619}.
\adsurl{2008ApJ...688L.115K}.
\end{barticle}
\endbibitem

\bibitem[\protect\citeauthoryear{{Leighton}}{1969}]{Leighton69}
\begin{barticle}
\bauthor{\bsnm{{Leighton}}, \binits{R.B.}}:
\byear{1969},
\batitle{{A Magneto-Kinematic Model of the Solar Cycle}}.
\bjtitle{\apj}
\bvolume{156},
\bfpage{1}.
\doiurl{https://doi.org/10.1086/149943}.
\adsurl{1969ApJ...156....1L}.
\end{barticle}
\endbibitem

\bibitem[\protect\citeauthoryear{Li}{2018}]{Li18}
\begin{barticle}
\bauthor{\bsnm{Li}, \binits{J.}}:
\byear{2018},
\batitle{A systematic study of hale and anti-hale sunspot physical parameters}.
\bjtitle{The Astrophysical Journal}
\bvolume{867}(\bissue{2}),
\bfpage{89}.
\burl{http://stacks.iop.org/0004-637X/867/i=2/a=89}.
\end{barticle}
\endbibitem

\bibitem[\protect\citeauthoryear{{Li} and {Ulrich}}{2012}]{Li12}
\begin{barticle}
\bauthor{\bsnm{{Li}}, \binits{J.}},
\bauthor{\bsnm{{Ulrich}}, \binits{R.K.}}:
\byear{2012},
\batitle{Long-term measurements of sunspot magnetic tilt angles}.
\bjtitle{The Astrophysical Journal}
\bvolume{758}(\bissue{2}),
\bfpage{115}.
\burl{http://stacks.iop.org/0004-637X/758/i=2/a=115}.
\end{barticle}
\endbibitem

\bibitem[\protect\citeauthoryear{{Lockwood} \textit{et~al.}}{2017}]{Lockwood17}
\begin{barticle}
\bauthor{\bsnm{{Lockwood}}, \binits{M.}},
\bauthor{\bsnm{{Owens}}, \binits{M.J.}},
\bauthor{\bsnm{{Imber}}, \binits{S.M.}},
\bauthor{\bsnm{{James}}, \binits{M.K.}},
\bauthor{\bsnm{{Bunce}}, \binits{E.J.}},
\bauthor{\bsnm{{Yeoman}}, \binits{T.K.}}:
\byear{2017},
\batitle{{Coronal and heliospheric magnetic flux circulation and its relation
  to open solar flux evolution}}.
\bjtitle{Journal of Geophysical Research (Space Physics)}
\bvolume{122},
\bfpage{5870}.
\doiurl{https://doi.org/10.1002/2016JA023644}.
\adsurl{2017JGRA..122.5870L}.
\end{barticle}
\endbibitem

\bibitem[\protect\citeauthoryear{{Makarov} and {Sivaraman}}{1986}]{Makarov86}
\begin{botherref}
\oauthor{\bsnm{{Makarov}}, \binits{V.I.}},
\oauthor{\bsnm{{Sivaraman}}, \binits{K.R.}}:
1986,
{Atlas of H-alpha synoptic charts for solar cycle 19 (1955 - 1964). Carrington
  solar rotations 1355 to 1486.}
\textit{Kodaikanal Observatory Bulletins}
\textbf{7}.
\adsurl{1986KodOB...7.....M}.
\end{botherref}
\endbibitem

\bibitem[\protect\citeauthoryear{{Makarov}, {Fatianov}, and
  {Sivaraman}}{1983}]{Makarov83}
\begin{barticle}
\bauthor{\bsnm{{Makarov}}, \binits{V.I.}},
\bauthor{\bsnm{{Fatianov}}, \binits{M.P.}},
\bauthor{\bsnm{{Sivaraman}}, \binits{K.R.}}:
\byear{1983},
\batitle{{Poleward migration of the magnetic neutral line and the reversal of
  the polar fields on the sun. I - Period 1945-1981}}.
\bjtitle{\solphys}
\bvolume{85},
\bfpage{215}.
\doiurl{https://doi.org/10.1007/BF00148649}.
\adsurl{1983SoPh...85..215M}.
\end{barticle}
\endbibitem

\bibitem[\protect\citeauthoryear{{McClintock} and
  {Norton}}{2013}]{McClintock13}
\begin{barticle}
\bauthor{\bsnm{{McClintock}}, \binits{B.H.}},
\bauthor{\bsnm{{Norton}}, \binits{A.A.}}:
\byear{2013},
\batitle{{Recovering Joy's Law as a Function of Solar Cycle, Hemisphere, and
  Longitude}}.
\bjtitle{\solphys}
\bvolume{287},
\bfpage{215}.
\doiurl{https://doi.org/10.1007/s11207-013-0338-0}.
\adsurl{2013SoPh..287..215M}.
\end{barticle}
\endbibitem

\bibitem[\protect\citeauthoryear{{McClintock} and
  {Norton}}{2016}]{McClintock16}
\begin{barticle}
\bauthor{\bsnm{{McClintock}}, \binits{B.H.}},
\bauthor{\bsnm{{Norton}}, \binits{A.A.}}:
\byear{2016},
\batitle{{Tilt Angle and Footpoint Separation of Small and Large Bipolar
  Sunspot Regions Observed with HMI}}.
\bjtitle{\apj}
\bvolume{818},
\bfpage{7}.
\doiurl{https://doi.org/10.3847/0004-637X/818/1/7}.
\adsurl{2016ApJ...818....7M}.
\end{barticle}
\endbibitem

\bibitem[\protect\citeauthoryear{{McIntosh}}{1979}]{McIntosh79}
\begin{botherref}
\oauthor{\bsnm{{McIntosh}}, \binits{P.S.}}:
1979,
{Annotated atlas of H-alpha synoptic charts for solar cycle 20 (1964-1974)
  Carrington solar rotations 1487-1616}.
\textit{NASA STI/Recon Technical Report N}
\textbf{79}.
\adsurl{1979STIN...7934149M}.
\end{botherref}
\endbibitem

\bibitem[\protect\citeauthoryear{{Mordvinov} and {Yazev}}{2014}]{Mordvinov14}
\begin{barticle}
\bauthor{\bsnm{{Mordvinov}}, \binits{A.V.}},
\bauthor{\bsnm{{Yazev}}, \binits{S.A.}}:
\byear{2014},
\batitle{{Reversals of the Sun's Polar Magnetic Fields in Relation to Activity
  Complexes and Coronal Holes}}.
\bjtitle{\solphys}
\bvolume{289},
\bfpage{1971}.
\doiurl{https://doi.org/10.1007/s11207-013-0456-8}.
\adsurl{2014SoPh..289.1971M}.
\end{barticle}
\endbibitem

\bibitem[\protect\citeauthoryear{{Mordvinov}, {Grigoryev}, and
  {Erofeev}}{2015}]{Mordvinov15}
\begin{barticle}
\bauthor{\bsnm{{Mordvinov}}, \binits{A.V.}},
\bauthor{\bsnm{{Grigoryev}}, \binits{V.M.}},
\bauthor{\bsnm{{Erofeev}}, \binits{D.V.}}:
\byear{2015},
\batitle{{Evolution of sunspot activity and inversion of the Sun's polar
  magnetic field in the current cycle}}.
\bjtitle{Advances in Space Research}
\bvolume{55},
\bfpage{2739}.
\doiurl{https://doi.org/10.1016/j.asr.2015.02.013}.
\adsurl{2015AdSpR..55.2739M}.
\end{barticle}
\endbibitem

\bibitem[\protect\citeauthoryear{{Mordvinov}
  \textit{et~al.}}{2016}]{Mordvinov16}
\begin{barticle}
\bauthor{\bsnm{{Mordvinov}}, \binits{A.}},
\bauthor{\bsnm{{Pevtsov}}, \binits{A.}},
\bauthor{\bsnm{{Bertello}}, \binits{L.}},
\bauthor{\bsnm{{Petri}}, \binits{G.}}:
\byear{2016},
\batitle{{The reversal of the Sun's magnetic field in cycle 24}}.
\bjtitle{Solar-Terrestrial Physics}
\bvolume{2}(\bissue{1}),
\bfpage{3}.
\doiurl{https://doi.org/10.12737/16356}.
\adsurl{2016STP.....2a...3M}.
\end{barticle}
\endbibitem

\bibitem[\protect\citeauthoryear{{Mu{\~n}oz-Jaramillo}
  \textit{et~al.}}{2013}]{Munoz13}
\begin{barticle}
\bauthor{\bsnm{{Mu{\~n}oz-Jaramillo}}, \binits{A.}},
\bauthor{\bsnm{{Dasi-Espuig}}, \binits{M.}},
\bauthor{\bsnm{{Balmaceda}}, \binits{L.A.}},
\bauthor{\bsnm{{DeLuca}}, \binits{E.E.}}:
\byear{2013},
\batitle{{Solar Cycle Propagation, Memory, and Prediction: Insights from a
  Century of Magnetic Proxies}}.
\bjtitle{\apjl}
\bvolume{767},
\bfpage{L25}.
\doiurl{https://doi.org/10.1088/2041-8205/767/2/L25}.
\adsurl{2013ApJ...767L..25M}.
\end{barticle}
\endbibitem

\bibitem[\protect\citeauthoryear{{Olemskoy} and
  {Kitchatinov}}{2013}]{Olemskoy13}
\begin{barticle}
\bauthor{\bsnm{{Olemskoy}}, \binits{S.V.}},
\bauthor{\bsnm{{Kitchatinov}}, \binits{L.L.}}:
\byear{2013},
\batitle{{Grand Minima and North-South Asymmetry of Solar Activity}}.
\bjtitle{\apj}
\bvolume{777},
\bfpage{71}.
\doiurl{https://doi.org/10.1088/0004-637X/777/1/71}.
\adsurl{2013ApJ...777...71O}.
\end{barticle}
\endbibitem

\bibitem[\protect\citeauthoryear{{Petrie} and {Ettinger}}{2017}]{Petrie17}
\begin{barticle}
\bauthor{\bsnm{{Petrie}}, \binits{G.}},
\bauthor{\bsnm{{Ettinger}}, \binits{S.}}:
\byear{2017},
\batitle{{Polar Field Reversals and Active Region Decay}}.
\bjtitle{\ssr}
\bvolume{210},
\bfpage{77}.
\doiurl{https://doi.org/10.1007/s11214-015-0189-0}.
\adsurl{2017SSRv..210...77P}.
\end{barticle}
\endbibitem

\bibitem[\protect\citeauthoryear{{Petrie}}{2015}]{Petrie15}
\begin{barticle}
\bauthor{\bsnm{{Petrie}}, \binits{G.J.D.}}:
\byear{2015},
\batitle{{Solar Magnetism in the Polar Regions}}.
\bjtitle{Living Reviews in Solar Physics}
\bvolume{12},
\bfpage{5}.
\doiurl{https://doi.org/10.1007/lrsp-2015-5}.
\adsurl{2015LRSP...12....5P}.
\end{barticle}
\endbibitem

\bibitem[\protect\citeauthoryear{{Pevtsov} \textit{et~al.}}{2014}]{Pevtsov14}
\begin{barticle}
\bauthor{\bsnm{{Pevtsov}}, \binits{A.A.}},
\bauthor{\bsnm{{Berger}}, \binits{M.A.}},
\bauthor{\bsnm{{Nindos}}, \binits{A.}},
\bauthor{\bsnm{{Norton}}, \binits{A.A.}},
\bauthor{\bsnm{{van Driel-Gesztelyi}}, \binits{L.}}:
\byear{2014},
\batitle{{Magnetic Helicity, Tilt, and Twist}}.
\bjtitle{\ssr}
\bvolume{186},
\bfpage{285}.
\doiurl{https://doi.org/10.1007/s11214-014-0082-2}.
\adsurl{2014SSRv..186..285P}.
\end{barticle}
\endbibitem

\bibitem[\protect\citeauthoryear{{Schrijver} and {Zwaan}}{2000}]{Schrijver00}
\begin{bbook}
\bauthor{\bsnm{{Schrijver}}, \binits{C.J.}},
\bauthor{\bsnm{{Zwaan}}, \binits{C.}}:
\byear{2000},
\bbtitle{{Solar and Stellar Magnetic Activity}}.
\adsurl{2000ssma.book.....S}.
\end{bbook}
\endbibitem

\bibitem[\protect\citeauthoryear{{Sheeley} and {Wang}}{2016}]{Sheeley16}
\begin{botherref}
\oauthor{\bsnm{{Sheeley}}, \binits{N.R.} \bsuffix{Jr.}},
\oauthor{\bsnm{{Wang}}, \binits{Y.-M.}}:
2016,
Bipolar mangetic regions determined from kitt peak vacuum telescope
  magnetograms.
\doiurl{https://doi.org/10.7910/DVN/TF6TY4}.
\url{https://doi.org/10.7910/DVN/TF6TY4}.
\end{botherref}
\endbibitem

\bibitem[\protect\citeauthoryear{{Snodgrass}, {Kress}, and
  {Wilson}}{2000}]{Snodgrass00}
\begin{barticle}
\bauthor{\bsnm{{Snodgrass}}, \binits{H.B.}},
\bauthor{\bsnm{{Kress}}, \binits{J.M.}},
\bauthor{\bsnm{{Wilson}}, \binits{P.R.}}:
\byear{2000},
\batitle{{Observations of the Polar Magnetic Fields During the Polarity
  Reversals of Cycle 22}}.
\bjtitle{\solphys}
\bvolume{191},
\bfpage{1}.
\doiurl{https://doi.org/10.1023/A:1005279508869}.
\adsurl{2000SoPh..191....1S}.
\end{barticle}
\endbibitem

\bibitem[\protect\citeauthoryear{{Sokoloff}, {Khlystova}, and
  {Abramenko}}{2015}]{Sokoloff15}
\begin{barticle}
\bauthor{\bsnm{{Sokoloff}}, \binits{D.}},
\bauthor{\bsnm{{Khlystova}}, \binits{A.}},
\bauthor{\bsnm{{Abramenko}}, \binits{V.}}:
\byear{2015},
\batitle{{Solar small-scale dynamo and polarity of sunspot groups}}.
\bjtitle{\mnras}
\bvolume{451},
\bfpage{1522}.
\doiurl{https://doi.org/10.1093/mnras/stv1036}.
\adsurl{2015MNRAS.451.1522S}.
\end{barticle}
\endbibitem

\bibitem[\protect\citeauthoryear{{Starck} and {Murtagh}}{2006}]{Starck06}
\begin{bbook}
\bauthor{\bsnm{{Starck}}, \binits{J.-L.}},
\bauthor{\bsnm{{Murtagh}}, \binits{F.}}:
\byear{2006},
\bbtitle{{Astronomical Image and Data Analysis}}.
\doiurl{https://doi.org/10.1007/978-3-540-33025-7}.
\adsurl{2006aida.book.....S}.
\end{bbook}
\endbibitem

\bibitem[\protect\citeauthoryear{{Stenflo} and {Kosovichev}}{2012}]{Stenflo12}
\begin{barticle}
\bauthor{\bsnm{{Stenflo}}, \binits{J.O.}},
\bauthor{\bsnm{{Kosovichev}}, \binits{A.G.}}:
\byear{2012},
\batitle{Bipolar magnetic regions on the sun: Global analysis of the soho/mdi
  data set}.
\bjtitle{The Astrophysical Journal}
\bvolume{745}(\bissue{2}),
\bfpage{129}.
\burl{http://stacks.iop.org/0004-637X/745/i=2/a=129}.
\end{barticle}
\endbibitem

\bibitem[\protect\citeauthoryear{{Sun} \textit{et~al.}}{2015}]{Sun15}
\begin{barticle}
\bauthor{\bsnm{{Sun}}, \binits{X.}},
\bauthor{\bsnm{{Hoeksema}}, \binits{J.T.}},
\bauthor{\bsnm{{Liu}}, \binits{Y.}},
\bauthor{\bsnm{{Zhao}}, \binits{J.}}:
\byear{2015},
\batitle{{On Polar Magnetic Field Reversal and Surface Flux Transport During
  Solar Cycle 24}}.
\bjtitle{\apj}
\bvolume{798},
\bfpage{114}.
\doiurl{https://doi.org/10.1088/0004-637X/798/2/114}.
\adsurl{2015ApJ...798..114S}.
\end{barticle}
\endbibitem

\bibitem[\protect\citeauthoryear{{Tlatova} \textit{et~al.}}{2018}]{Tlatova18}
\begin{barticle}
\bauthor{\bsnm{{Tlatova}}, \binits{K.}},
\bauthor{\bsnm{{Tlatov}}, \binits{A.}},
\bauthor{\bsnm{{Pevtsov}}, \binits{A.}},
\bauthor{\bsnm{{Mursula}}, \binits{K.}},
\bauthor{\bsnm{{Vasil'eva}}, \binits{V.}},
\bauthor{\bsnm{{Heikkinen}}, \binits{E.}},
\bauthor{\bsnm{{Bertello}}, \binits{L.}},
\bauthor{\bsnm{{Pevtsov}}, \binits{A.}},
\bauthor{\bsnm{{Virtanen}}, \binits{I.}},
\bauthor{\bsnm{{Karachik}}, \binits{N.}}:
\byear{2018},
\batitle{{Tilt of Sunspot Bipoles in Solar Cycles 15 to 24}}.
\bjtitle{\solphys}
\bvolume{293},
\bfpage{118}.
\doiurl{https://doi.org/10.1007/s11207-018-1337-y}.
\adsurl{2018SoPh..293..118T}.
\end{barticle}
\endbibitem

\bibitem[\protect\citeauthoryear{Upton and Hathaway}{2014}]{Upton14}
\begin{barticle}
\bauthor{\bsnm{Upton}, \binits{L.}},
\bauthor{\bsnm{Hathaway}, \binits{D.H.}}:
\byear{2014},
\batitle{Effects of meridional flow variations on solar cycles 23 and 24}.
\bjtitle{The Astrophysical Journal}
\bvolume{792}(\bissue{2}),
\bfpage{142}.
\burl{http://stacks.iop.org/0004-637X/792/i=2/a=142}.
\end{barticle}
\endbibitem

\bibitem[\protect\citeauthoryear{Wang and Sheeley}{1989}]{Wang89}
\begin{barticle}
\bauthor{\bsnm{Wang}, \binits{Y.-M.}},
\bauthor{\bsnm{Sheeley}, \binits{N.R.}}:
\byear{1989},
\batitle{Average properties of bipolar magnetic regions during sunspot cycle
  21}.
\bjtitle{Solar Physics}
\bvolume{124}(\bissue{1}),
\bfpage{81}.
\doiurl{https://doi.org/10.1007/BF00146521}.
\burl{https://doi.org/10.1007/BF00146521}.
\end{barticle}
\endbibitem

\bibitem[\protect\citeauthoryear{{Wang} and {Sheeley}}{1991}]{Wang91}
\begin{barticle}
\bauthor{\bsnm{{Wang}}, \binits{Y.-M.}},
\bauthor{\bsnm{{Sheeley}}, \binits{N.R.} \bsuffix{Jr.}}:
\byear{1991},
\batitle{{Magnetic flux transport and the sun's dipole moment - New twists to
  the Babcock-Leighton model}}.
\bjtitle{\apj}
\bvolume{375},
\bfpage{761}.
\doiurl{https://doi.org/10.1086/170240}.
\adsurl{1991ApJ...375..761W}.
\end{barticle}
\endbibitem

\bibitem[\protect\citeauthoryear{{Wang} and {Sheeley}}{2004}]{Wang04}
\begin{barticle}
\bauthor{\bsnm{{Wang}}, \binits{Y.-M.}},
\bauthor{\bsnm{{Sheeley}}, \binits{N.R.} \bsuffix{Jr.}}:
\byear{2004},
\batitle{Footpoint switching and the evolution of coronal holes}.
\bjtitle{The Astrophysical Journal}
\bvolume{612}(\bissue{2}),
\bfpage{1196}.
\burl{http://stacks.iop.org/0004-637X/612/i=2/a=1196}.
\end{barticle}
\endbibitem

\bibitem[\protect\citeauthoryear{{Webb} \textit{et~al.}}{2017}]{Webb17}
\begin{botherref}
\oauthor{\bsnm{{Webb}}, \binits{D.F.}},
\oauthor{\bsnm{{Gibson}}, \binits{S.E.}},
\oauthor{\bsnm{{Hewins}}, \binits{I.}},
\oauthor{\bsnm{{McFadden}}, \binits{R.}},
\oauthor{\bsnm{{Emery}}, \binits{B.A.}},
\oauthor{\bsnm{{Malanushenko}}, \binits{A.V.}}:
2017,
{Studies of Global Solar Magnetic Field Patterns Using a Newly Digitized
  Archive}.
\textit{AGU Fall Meeting Abstracts}.
\adsurl{2017AGUFMSH54A..01H}.
\end{botherref}
\endbibitem

\bibitem[\protect\citeauthoryear{{Yeates}, {Baker}, and {van
  Driel-Gesztelyi}}{2015}]{Yeates15}
\begin{barticle}
\bauthor{\bsnm{{Yeates}}, \binits{A.R.}},
\bauthor{\bsnm{{Baker}}, \binits{D.}},
\bauthor{\bsnm{{van Driel-Gesztelyi}}, \binits{L.}}:
\byear{2015},
\batitle{{Source of a Prominent Poleward Surge During Solar Cycle 24}}.
\bjtitle{\solphys}
\bvolume{290},
\bfpage{3189}.
\doiurl{https://doi.org/10.1007/s11207-015-0660-9}.
\adsurl{2015SoPh..290.3189Y}.
\end{barticle}
\endbibitem

\end{thebibliography}
     % name your Bibtex file containing your references (.bib)

\end{article}

\end{document}